*Chapter 1*

# LIVING WITH RADICAL UNCERTAINTY: THE EXEMPLARY CASE OF FOLDING PROTEIN

*Ignazio Licata**

ISEM, Institute for Scientific Methodology, Palermo

## ABSTRACT

Laplace's demon still makes strong impact on contemporary science, in spite of the fact that Logical Mathematics outcomes, Quantum Physics advent and more recently Complexity Science have pointed out the crucial role of uncertainty in the World's descriptions. We focus here on the typical problem of folding protein as an example of uncertainty, radical emergence and a guide to the "simple" principles for studying complex systems.

## 1. "ZIP FILING THE WORLD"

A ghost roams science, a sort of hidden paradigm - never formally enunciated – which steers the way how the whole scientific activity is conceived. We mean the completely-computable-world idea whose conceptual foundation lies reductionism. If we had to trace back the manifesto of such a conceptual tension, the famous Laplace's excerpt from his *Essai Philosophique sur les Probabilités* (1814) is exemplary:

> "We must therefore regard the present state of universe as the effect of its preceding state and as the cause of one which is to follow. An intelligence which in a single instant could know all the forces which animate the natural world, and the respective situations of all the beings that made it up, could, provided it was vast enough to make an analysis of all the data so supplied, be able to produce a single formula which specified all the movements in the universe from those of the largest bodies in the universe to those of the lightest atom. For such intelligence, nothing would be "uncertain", and the nature, like the past, would be present before its eyes."

---

* Ignazio.licata@ejtp.info

Although the very nature of Laplace's observer – with its ability to measure the initial condition of all the particles in the Universe, to insert them in the Newtonian equations and so to calculate any trajectory – patently appears as extremely speculative, the interesting side is the implicit assuming that such a program is unfeasible, and yet *theoretically possible*! In other words, there is no contradiction in *thinking it out*, and it is in consonance with our knowledge of the physical world.

Still nowadays, contrary to what it is argued, neither Quantum Physics nor non-linear dynamics have swayed such conception. In Quantum Mechanics, for example, the Schrödinger's equation characterizes an evolutionary dynamics belonging to **U**-type – according to the well-known Penrose classification – which is perfectly deterministic (besides, from the structural viewpoint, the non-relativistic quantum physics' key equation is similar to a diffusion equation and formally connected to the Hamilton-Jacobi equation), whereas the **R** processes, which the probabilistic interpretation is centered on and related to the state vector collapse , can always be regarded as the outcome of an "hidden determinism" limiting any attempt to read them classically, but has no radical incidence on the idea of a complete calculability of the World a la Laplace (Allori & Zanghì., 2004).

The same goes for non-linear dynamics: the sensitivity to initial conditions and the long-term unpredictability does not remove local determinism; in fact they make the chaotic dynamics as the ideal sample of computational emergence, algorithmically compressible in few simple formulas (see Gleick, 2008; Chaitin, 2007). It is not by chance the close analogy between the halting problem in Turing computation theory and the deterministic chaotic systems: in both of them the final status is unpredictable, but it can be followed step-by-step. Even the structural instability that was studied by Pontryagin, Andronov and Peixoto in mathematics and whose equivalent in Physics are the Haken and Prigogine dissipative systems – where, starting from a situation of instability, a system can choose infinite equilibrium states – does not prevent a "global" forecast about the asymptotic state of the dynamical evolution (Thom, 1994; Prigogine & Nicolis, 1989; Haken, 2006).

Despite its Newtonian roots has grown weaker and weaker, the Laplace-inspired computability of the World has not yet been completely undermined by the Quantum Physics and non-linear dynamics advent. The mythology of "Everything Theories" is grounded on such line of thought (Barrow, 2008), where the key idea is "Zip-filing the World" into a fistful of formulas describing the fundamental interactions between a restricted group of "fundamental objects". Two of the main reductionism meanings fall within such mythology. The more patent one is related to the crucial – nomological, we dare say - role of the "World's bricks"; the other and subtler one suggests that the World can be described by a theoretic chain of the kind $T_1 \langle T_2 \langle ... T_f$, where the $T_i$ are the description levels and the symbol "$\langle$" means "physically weaker than", and consequently each level can be derived by the "final theory" $T_f$.

We find interesting to point out that even in a so deeply changed cultural context the idea of the World ultimate computability still largely characterizes the conception of scientific activity as well as the scientific method role, both regarded as the asymptotic approaching to the "reality's ultimate structure". One of the principal consequences of such way of thinking lies in considering uncertainty not only as bad, but a worthless bad feature, a mere practical drag to knowledge and located just in some remote and radical quantum zone ruled by Heisenberg Indetermination Principle. Any science where uncertainty could not be tamed so easily was downgraded to the status of "imperfect science", unfit for the Physics-oriented programme, as it happened with Biology, Cognitive and Socio-Economic Sciences.

Paul Anderson in his famous paper-manifesto "More is Different" (1972) radically and subtly criticized such programme, a criticism later developed by Laughlin and Pines (Laughlin & Pines, 2000; Laughlin, 2006). The central idea is simple: the universality of the collective behaviors - such as phase transitions – is *compatible* with the system's constituents, but not deducible from the properties of the elementary "bricks". Thus, reductionism simply does not work with such kind of systems, which, after all, are the great majority of the *interesting* systems and stays in that zone called the "Middle Way" (Laughlin et al., 2000) that is located between Particle Physics and Cosmology. In "middle-way" systems, evolution is strongly connected to dynamical coupling with environment and depends on the structural history of the system. The "Everything Theory" for this kind of systems is impossible, just because any interactions between *every single* system and the environment at each instant should be taken into account. That is to state the best dynamical description of the system is the system itself! Contrary to the Laplace hypothesis, the observer is "immersed in the World" and has to make its descriptive choices critically depending on its inter-relation with the system (Licata, 2008).

So, we have to come to terms with the idea that a mathematical model of a complex system is just a "picture" of a single side of the system that has been taken from one of the several possible perspectives among the organizational behaviors. Does this kind of dynamical usage of models fall under the scheme of the above-examined theoretical chain? The answer is no! Uncertainty and incommensurability between a description and another one are here natural elements of the knowledge process and are placed in the interstices of any theoretical description. In fact, a model is not a mere description of a space-time range (for ex. from the smallest to the biggest), but it is fixed by the observer's *aim* and its goals; thus models aimed at grasping a specific behavior of the system are intrinsically affected by uncertainty as for the other aspects and they cannot be easily tuned within a single key (Minati, 2008; Licata, 2008).

We will show here that Folding Protein is a perfect example of a "middle-way" process as well as a precious guide to the understanding of both complex systems and the radical elements of uncertainty which are connected to their description.

## 2. THE FOLDING PROBLEM

Proteins are linear heteropolymers made of a non-periodic chain of amino acids connected by a covalent bond. The vital task for "life mechanism" is carried out by folding process, when a protein in solution folds into a three-dimensional structure which must be self-consistent with the solvent. Proteins catalyze the chemical reactions necessary to life (enzymes), build the biological structure in the strict sense and individuate antibodies. These macromolecules, at the edge between Chemistry and Biology, carry a "biological signature" containing all the significant elements of the authentic biological complexity. The problem a protein has to solve is that of finding a spatial soluble configuration so as to carry out its functions without going out of the solution. In this condition, the protein can carry out its informational role, which depends on a great deal of boundary conditions related to environment where it is immersed (solvent, presence of other molecules, PH, ionic strength and so on) (see Whitford, 2005; Giuliani and Zbilut, 2008).

In spite of the huge proposed classifications there do not exist identical folding configurations, each one depends on the responses of "actors into play" to the specific context (as for proteins these actors are well-known: hydrophobic bond, hydrogen bond, van der Waals forces, electrostatic interaction).

Such singular coupling with environment is a general feature of any complex system and is valid for a protein as well as a social group or a factory. Uncertainty is the rule there. And it provides for a "minimal" definition of complex system: *a complex system is a system showing locally an unpredictable behavior, which could not be zip-filed in a single formal model.* In highly logical open systems (which not only exchange matter/energy, but modify their arrangement too) it is impossible to distinguish "*in vitro*" behavior (closed systems describable by a single formal model) from *"in vivo"* behavior. It follows what we call the first principle of complex systems or logical openness: *complex systems are open, context-sensitive systems.*

The final configuration of a crystal is unique, whereas the protein's dynamic path toward its three-folded state is quite "rugged", full of false minima and metastable states (see Figure 1).

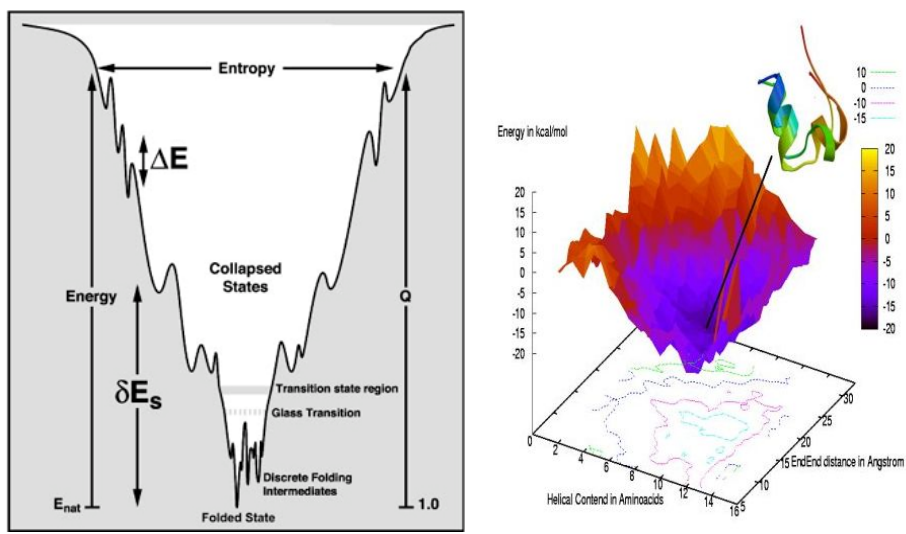

Figure 1. Folding funnel in rugged landscape.

Also such feature lends itself to a generalization valid for all complex systems and is strictly related to unpredictability and uncertainty, the second "simple" principle is the Principle of Indifference: *a complex system exhibits several different behaviors equivalent from energetic viewpoint and thus impossible to classify into a hierarchical order, <u>not even a probabilistic one.</u>*

The reason why we have underlined the last sentence is to point out we are dealing with a more radical unpredictability than the quantum or non-linear ones and at the same its nature is closer to our everyday experience. It is a "sliding-doors"-type of indifference, the system can choose different possible "histories", each one following a completely different "destiny". In Quantum Physics, within given situations, it is possible to assign a probability weight as well as in non-linear dynamics of the simplest models everything depends in continuous way on initial conditions. In the above-described cases instead not only it is impossible any *a priori* evaluation, but the system's dynamic history is really far from following the "domestic" rules of differential calculus to such an extent that we need a mathematics taking into account singularities, terminal points and environmental noise rather than trajectories (Non-Lipshitz dynamics, see Zbilut, 2004). Thus it follows the third principle, or "it is easier to observe it": *the path that a system follows towards its final state is decisive to define the state itself. Complex systems can only be narrated by 'consequent stories' and, not by a priori predefined ones*. It has to be pointed out how such statements which appear so weird if related to the Physics' traditional problems – where boundary and initial conditions are very different from the "law" ruling the phenomenon – become absolutely obvious when we move to different field such as cognitive processes. That is why both Artificial Intelligence and the research for "algorithmic laws of thinking" have failed (Licata, 2008). Thus, the fourth principle deals with the structure-function relationship, *inextricableness of the structure from a complex system dynamics: a system is its own history!*

## 3. RADICAL EMERGENCE: *PYROCOCCUS FURIOSUS*

The general features of complex systems provide them with "flexibility" and adaptedness enabling life and cognition and show the deep link between unpredictability, uncertainty and emergence. That is the principle of "surprise": *complex systems exhibit radical emergence properties*. "Radical" is here referred to the appearing of properties which cannot be deduced from a predefined model of the system as responses to the specific situation of coupling with the environment. An Universe a la Laplace can be totally assimilated to a Turing machine, where the observer itself play the role of "event recorder": many things happen, but all of them fall under the "cosmic code" of fundamental laws. In complex systems, instead, is the *single process* that counts, and *how it occurs*, "laws" are just the stable elements within a very tangled dynamic frame, where surprises not deducible from a single model emerge.

The study of thermophilic proteins, able to live without denaturing at very high temperatures, is extremely interesting. That is the case of *Pyrococcus Furiosus,* an endemic archaea bacterium located in Pozzuoli sulphur mines at about 90° C (194° F), a temperature usually causing mesophilic proteins to loose their tertiary structure. A purely physico-chemical analysis cannot find an explanation for the peculiarities of themophilic proteins, as we has already said the "bricks" are the same and the forces into play, too. By studying "chimeras", artificial proteins made of both thermophilic and mesophilic sequences, it has been shown that thermophily is a global feature of a system: a protein is either thermophilic or not-thermophilic, its lifetime does not depend on how the sequences are recombined!

Here it is an exemplary case of emergent property, compatible with structure, sequence, elements and forces into play, but not deducible from them all. New information about a system spur us to use new methodological approaches and update the model which, in turn, is a cognitive emergence caused by the phenomenon under consideration.

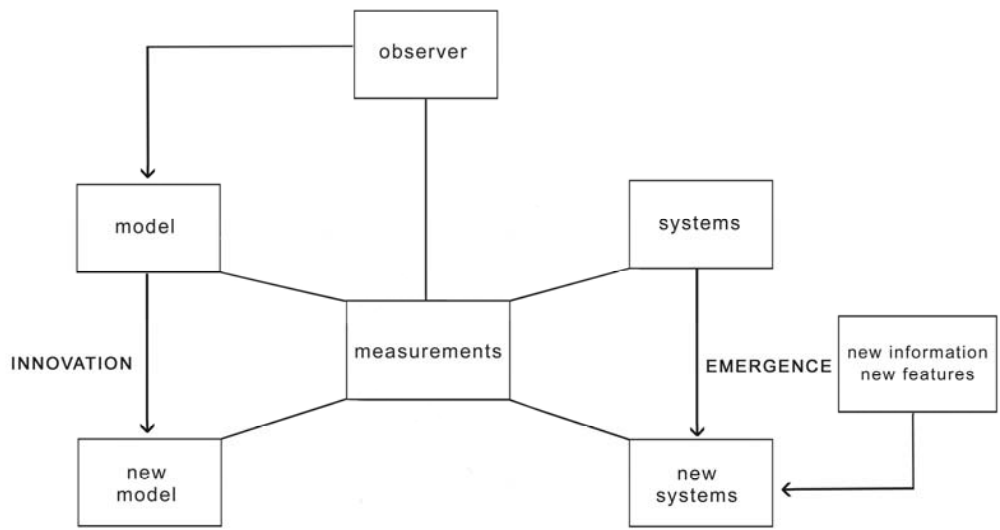

Figure 2. model updating under emergence.

A significant clue can be found in a statistical analysis technique called Recurrence Quantification Analysis (RQA) and developed by Webber and Zbilut in order to study complex systems such as the biological or financial ones, and so on (Zbilut and Webber, 1992, 1994). As its name suggests such analysis is based on recurrence graphs where the elements of data sequence are plotted in the same point if they indicates a similar position in the phase space. Practically, if the x-axis value and the y-axis one are very close (for two-dimensional case) – it is given for granted an impossible to remove structural uncertainty, so RQA is ideal for studying such systems – they individuate the same point. Such phase portraits are really illuminating. In the case of proteins, the recurrence between adjacent sequences is studied, the outcome can be found in Figure 3.

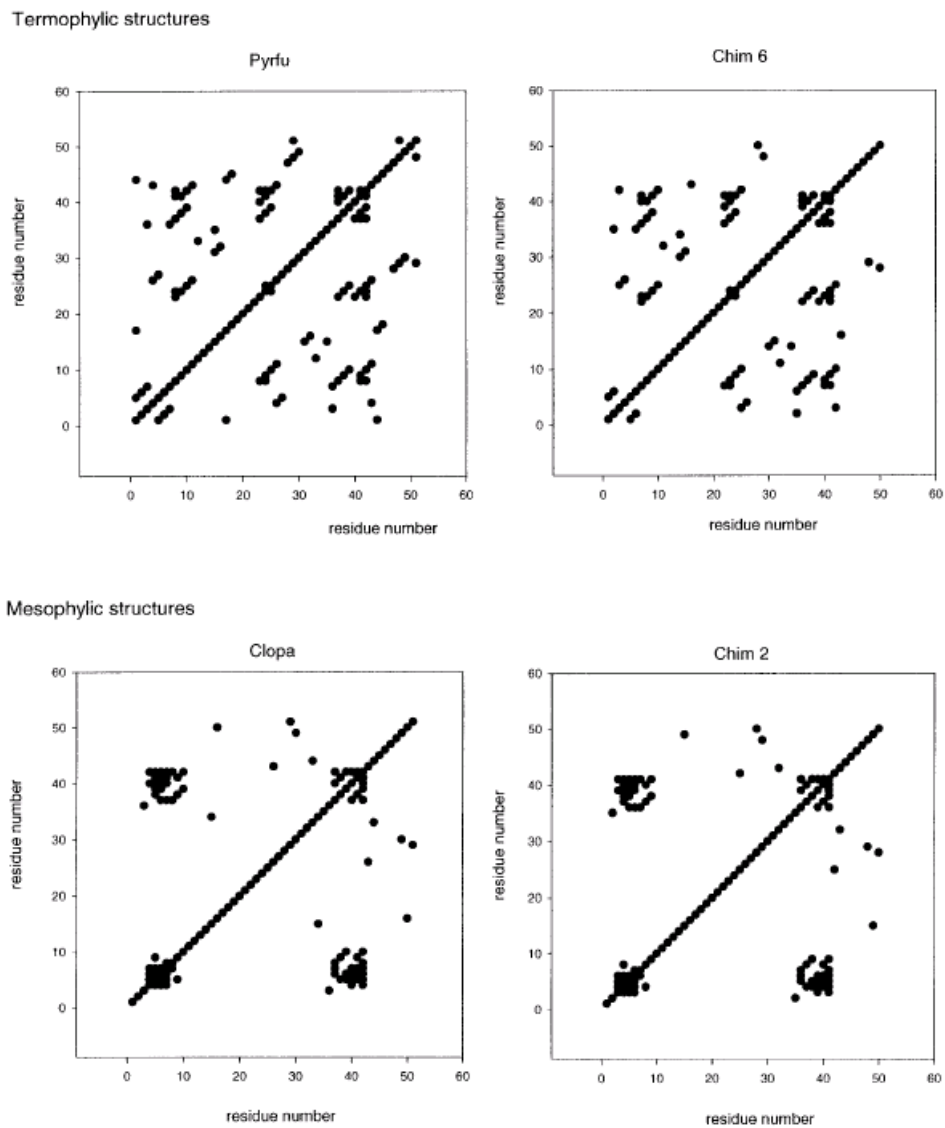

From Giuliani et al., 2000.

Figure 3. Comparing of phase portraits by RQA between thermophilic and mesophilic proteins (left) and two chimeras ( right).

The key element is the relation between hydrophilic and hydrophobic sequences, that is to say the "dialogue" between protein and its solvent (water).

In thermophilic proteins, the hydrophobicity patterns are widely distributed along lines parallel to the main diagonal, whereas they are more clustered in mesophilic ones. This thing directly depends on the Principle of Indifference – many solutions energetically equivalent – and has to be interpreted as a greater flexibility of the structure, elements and "architectural" project being the same!That also explains because there do not exist intermediate possibilities: under a given level of flexibility, the protein "collapses" and is unable to carry out its function.

What kind of "explanation" is this? Not a deductive one, because there are neither a formal model nor the analysis of the constituents to reveal it, but rather a "global" evidence which is derived *a posteriori* (it's easier to observe it) when we use methodological tools able to grasp these dynamical features of the system. That is the essential peculiarity of authentic emergent processes; they cannot be brought back to a specific local "cause", but rather to a set of collective conditions which allow its occurring. In the case of thermophily, such conditions can be found in the extension of the distribution of the possible configurations and in the transition speed-rate from a configuration to another one as the response to the high thermal excitation of the external environment. All that can happen by a particular architectural expedient related to hydrophobic sequences and reminds the spin-glasses' essential topics in Physics, where the collective characteristics of a system are deeply connected to the complex dynamic arrangement of many local equilibrium states.

## 4. UNCERTAINTY'S FECUNDITY: A SYSTEMIC CONCLUSION

This short *excursus* in Folding Protein has lead us towards some key principles that cal be applied to any complex system, from Biology to socio-economic systems. Why are they so simple? Should not we expect a complexity science as extremely complicated as the objects it studies? "Simplicity" depends on the fact that the more the collective behaviors matter, *the less* the microscopic detailed questions *are important*. Besides, it is a seeming simplicity, because it requires explanation styles and analysis methods completely different from those used to study simple systems which can be easily "enclosed" within an analytic solution. From a systemic viewpoint, simple systems lend themselves to be described by a model in which it is always possible to individuate a distinct border between system and environment, easily schematizing the inter-relations into play. In the great majority of the cases such procedure is the equivalent of making a "toy-model" and "sweeping complexity under the carpet". There does not exist a reassuring fixed "border" between a system and its environment, and even the identification of the elementary constituents changes in accordance with collective dynamics. Reductionism is a strategy that can bear fruit and has productively led the scientific explanation style for about three centuries, but it cannot be applied in any situation. Tackling complexity means to accept that radical uncertainty all of us experience in the plurality of the possible choices in our life, and which is somewhat subtler and more pervasive than the quantum one, which is fixed by Heisenberg principle.

We could be tempted by proposing again the Laplace's hypothesis: if such intelligence able to take into account each object in the Universe, each system-environment coupling did exist, could it overcome uncertainty? We suggested good reasons which make clear that assuming such kind of observer does not make any sense. "We get no the God's eye" is what Tibor Vamos (Vamos, 1993) efficaciously wrote; a post-Hegelian argumentation about an observer immersed into the World it is observing can e found in Breuer (Breuer, 1995). And yet the problem of Laplace absolute observer can provide an important systemic cue. In nature there do not exist only objects, but also the objects' behaviors which cannot be observed when we only focus on elementary constituents, even admitting that these ones can always be individuated without ambiguity. So the "universal" observer should not only be able to take into account every single behavior but also the myriads of collective behaviors in which the object can be involved at the same time! That is nothing but stating that the best "narration" of the World is the natural processes' evolution itself. Scientific observers, instead, are always situated, and uncertainty as well as the limits of their models are a strong spur for new explorations and perspectives.

## ACKNOWLEDGMENTS

The author owes a lot to Alessandro Giuliani for unveiling the fascinating world of proteins. This paper is dedicated to my friend Joe Zbilut memory (1948 – 2009).